\documentstyle[pre,aps]{revtex}

\newcommand{\sr}{\stackrel{\rightarrow}{r}}
\newcommand{\be}{\begin{equation}}
\newcommand{\ee}{\end{equation}}
\newcommand{\prt}{\partial}
\newcommand{\ra}{\rightarrow}

\begin{document}

\draft

\title{Negative Electric Current in Semiconductors} 
\author{V.I. Yukalov and E.P. Yukalova} 
\address{International Centre of Condensed Matter Physics \\
University of Brasilia CP 04513, Brasilia DF 70919--970, Brazil \\
and \\
Bogolubov Laboratory of Theoretical Physics \\
Joint Institute for Nuclear Research, Dubna 141980, Russia}

\maketitle

\begin{abstract}

Electric current in semiconductors is considered as a function of time. 
Nonequilibrium fluctuations of the current are studied for strongly 
nonuniform samples. It is demonstrated that at special conditions the total 
electric current through a device can exhibit a negative fluctuation, 
when the current is directed against the applied voltage.

\end{abstract}

\pacs{72.20.--i, 05.60.+w}

{\bf keywords:} transport equations, nonuniform semiconductors, 
electric--current fluctuations

\vspace{1cm}

The key to the understanding of the physical properties of semiconductor 
devices lies in the thorough analysis of equations modelling their 
behaviour [1]. The equations describing semiconductor transport 
characteristics are quite complicated, so that usually only stationary 
regimes can be satisfactory analysed. The situation becomes even more 
complicated when the distribution of charge carriers is essentially 
nonuniform. At the same time, strongly nonequilibrium and nonuniform 
semiconductors can display peculiar features that cannot appear in 
stationary and uniform samples [2--4].

In this communication we study nonequilibrium transport properties of 
semiconductor samples with a strongly nonuniform distribution of carrier 
densities. We show that at some special conditions there can appear a 
negative electric current, that is, the current directed against the 
applied voltage. The occurrence of this negative current is a transient 
effect that can be realized at the initial stage of a nonequilibrium 
process, while the carrier densities are yet essentially nonuniform. As 
soon as the charge distribution becomes more uniform, the current returns 
to normal. Thus, the negative--current effect is displayed as a kind of 
fluctuation happening because of strong nonuniformity of carriers in a 
nonequilibrium semiconductor.

The charge carriers are characterized by the densities 
$\rho_i=\rho_i(\sr,t)$, with $i=1,2$, which are functions of the space 
variable $\sr$ and of time $t$. Let $\rho_1>0$ and $\rho_2<0$. Carrier 
transport is commonly described in terms of the semiclassical 
drift--diffusion approximation which is the basis of semiconductor device 
models [1]. This approach rests on the continuity equations
\be
\frac{\prt\rho_i}{\prt t} + \stackrel{\ra}{\nabla}\stackrel{\ra}{j}_i +
\gamma_i\rho_i = 0
\ee
and on the Maxwell equation
\be
\varepsilon\stackrel{\ra}{\nabla}\stackrel{\ra}{E} = 4\pi(\rho_1+\rho_2) ,
\ee
with the electric--current density
\be
\stackrel{\ra}{j}_i=\mu_i\rho_i\stackrel{\ra}{E} -
D_i\stackrel{\ra}{\nabla}\rho_i ,
\ee
where $\gamma_i$ is a relaxation width, $\varepsilon$ is the dielectric
permittivity, $\mu_1>0$ and $\mu_2<0$ are the carrier mobilities, and $D_i$
is a diffusion coefficient.

Adding to the sum of the drift--diffusion currents in eq.(3) the displacement
current, one gets the total current density
\be
\stackrel{\ra}{j}_{tot} =\stackrel{\ra}{j}_1 +\stackrel{\ra}{j}_2
+\frac{\varepsilon}{4\pi}\frac{\prt\stackrel{\ra}{E}}{\prt t} .
\ee
The total current across a device is given by the integral
\be
\stackrel{\ra}{J}(t) =\int\stackrel{\ra}{j}_{tot}(\sr,t)d\sr ,
\ee
with the integration over the considered sample.

Consider a plane device of the width $L$ and area $A$, being biased with 
an external voltage $V_0$. Then, instead of the vector $\sr$, we have one 
space variable $x\in[0,L]$. Define the effective transient time
$$ \tau_0\equiv\frac{L^2}{\mu V_0} , \qquad \mu 
\equiv\frac{1}{2}(\mu_1 + |\mu_2|) , $$
where $\mu$ is the average mobility. It is convenient to scale physical 
characteristics so that to deal with dimensionless quantities. To this 
end, we shall measure in what follows the space variable $x$ in units of 
$L$; time $t$, in units of $\tau_0$; and other physical quantities, in 
the corresponding
units written below:
$$ \rho_0\equiv \frac{Q_0}{AL} , \qquad Q_0\equiv \varepsilon AE_0 , \qquad
E_0\equiv\frac{V_0}{L} , $$
$$ \gamma_0\equiv\frac{1}{\tau_0} , \qquad D_0 \equiv \mu V_0 , \qquad
J_0\equiv\frac{Q_0L}{\tau_0} . $$

For the plane geometry considered, eq.(1) reduces to
\be
\frac{\prt\rho_i}{\prt t} +\mu_i\frac{\prt}{\prt x}(\rho_iE) -
D_i\frac{\prt^2\rho_i}{\prt x^2} +\gamma_i\rho_i = 0
\ee
and eq.(2) reads
\be
\frac{\prt E}{\prt x} = 4\pi(\rho_1 +\rho_2) , 
\ee
where $0<x<1$ and $t>0$.

Assume that at initial time the sample is prepared with a nonuniform 
distribution of charge carriers
\be
\rho_i(x,0) = f_i(x) .
\ee 
There exist several ways of forming samples with nonuniform carrier 
densities. This can be done, e.g., by means of special procedures of 
growing a semiconductor with doped layers incorporated during the growth [1].
Another way is to subject the sample to the influence of irradiation by 
particle or laser beams [3].

Let the considered device be biased with an externally applied voltage 
$V_0$, which, in the dimensionless units accepted, implies the condition
\be
\int_0^1 E(x,t)dx = 1 .
\ee

For the electric field we may derive from eq.(7), with condition (9), the 
functional
\be
E(x,t) = 1+4\pi\left [ Q(x,t) -\int_0^1 Q(x,t)dx\right ]
\ee
of the charge densities entering through the notation
$$ Q(x,t) =\int_0^x\left [\rho_1(x',t) +\rho_2(x',t)\right ]dx' . $$

The total electric current (5) across the plane device can be written as
\be
J(t) =\int_0^1[ j_1(x,t) + j_2(x,t)]dx ,
\ee
where condition (9) is taken into account and
$$ j_i =\mu_i\rho_i E - D_i\frac{\prt\rho_i}{\prt x} . $$
From eq. (11) we find
$$ J(t) =\int_0^1\left [ \mu_1\rho_1(x,t) +\mu_2\rho_2(x,t)\right ] 
E(x,t)dx + $$
\be 
+ D_1\left [\rho_1(0,t) -\rho_2(1,t)\right ] + 
D_2\left [\rho_2(0,t) - \rho_2(1,t)\right ] .
\ee
To understand the time behaviour of the electric current, we need to 
analyse expression (12), in which $\rho_i$ and $E$ are defined by eqs.(6) 
and (7), with the initial condition (8) and the voltage condition 
(9). Such an analysis can be done resorting to computer calculations. 
However, before doing this, it is very useful to derive an approximate 
analytical solution clarifying the physics of processes.

An approximate solution to the system of eqs.(6) and (7) can be obtained 
employing the method of scale separation [5-7] which is a generalization 
of the Krylov--Bogolubov averaging method [8] and the of guiding--centre 
approach [9].

The first step in the method of scale separation [5-7] is the 
classification of solutions onto fast and slow. In the present case, this 
is done as follows. Notice that the electric field is expressed as the 
functional (10) containing the carrier densities integrated over space. 
Thence, $E$ should vary in space slower than $\rho_i$. On the other hand, 
the voltage condition (9) shows that the electric field, being averaged 
over space, does not depend on time. This means that $E$ can be regarded
as a slow function in time. Therefore, the electric field can be classified 
as a slow solution, with respect to both space and time, as compared to 
the carrier densities. This allows us to consider eq.(6) for a fast 
solution $\rho_i$ treating there $E$ as a space--time quasi--integral. 
Solving eq.(6), the found $\rho_i$ is to be substituted into functional 
(10), which results in an equation for $E$ to be solved iteratively.

When solving eq.(6), we continue $\rho_i$ outside the region $0<x<1$ by 
defining $\rho_i$ as zero for $x<0$ and $x>1$. As a result we find
\be
\rho_i(x,t) =\int_{-\infty}^{+\infty} G_i(x-x',t)f_i(x')dx' ,
\ee
where the Green function is
$$ G_i(x,t) =\frac{1}{2\sqrt{\pi D_it}}\exp\left\{ -
\frac{(x-\mu_iEt)^2}{4D_it}-\gamma_it\right\} . $$

As an initial condition in eq.(8), it is reasonable to accept the physically
realistic case of the Gaussian distribution
\be
f_i(x) =\frac{Q_i}{Z_i}\exp\left\{ -\frac{(x-a_i)^2}{2b_i}\right\} 
\ee
centered at the point $a_i\in(0,1)$ and in which
$$ Q_i=\int_0^1f_i(x)dx , \qquad
Z_i =\int_0^1\exp\left\{ -\frac{(x-a_i)^2}{2b_i}\right\} dx . $$
Then, from eq. (13) we obtain
\be
\rho_i(x,t) =\frac{Q_ib_i}{Z_i\sqrt{b_i^2 +2D_it}}\exp\left\{ -
\frac{(x-\mu_iEt-a_i)^2}{2b_i^2 + 4D_it}-\gamma_it\right\} .
\ee 
As is evident, at $t=0$ solution (15) reduces to the initial condition 
(14). And at $t\ra\infty$, we have
\be
\lim_{t\ra\infty}\rho_i(x,t) = 0 , \qquad
\lim_{t\ra\infty} E(x,t) = 1 .
\ee

Now, we have in hands enough information to try to find out whether a negative
electric current could really appear, that is, an electric current directed 
against the applied voltage $V_0$. Take, for concreteness, that $V_0>0$. Then 
we need to find when $J(t)<0$.

First thing, we immediately may notice, is that if $\rho_i(x,t)$ is uniform in
space then eq. (12) tells us that the electric current is a positively defined
quantity $J(t)=\mu_1\rho_1+\mu_2\rho_2>0$. So, if the electric current 
has a chance to become negative, this can happen only when $\rho_i(x,t)$ 
is nonuniform in space.

From solution (15) it follows that even if the carrier densities are 
nonuniform at the initial time, nevertheless, they tend to become uniform 
with time, changing to zero, as $t\ra\infty$. Consequently, if a negative 
electric current could arise, this might happen only at the initial stage 
when $t\ll 1$ and $\rho_i(x,t)$ are yet essentially nonuniform. Thus, a 
negative electric current, if any, can occur only as a principally 
transient effect at time $t\ll 1$; and the space nonuniformity of the 
initial carrier density $\rho_i(x,0)=f_i(x)$ is a necessary condition for 
this effect.

Any initially nonuniform density, according to form (15), spreads with 
time because of diffusion and relaxation effects. However, these 
processes need some time to smooth $\rho_i$. Hence, there always can be 
found such a time $t\ll 1$ when $\rho_i$ is yet nonuniform. But diffusion 
and relaxation shorten the time of a negative--current fluctuation, if 
such arises. Therefore, the conditions favoring the appearance of the 
noticeable negative--current effect are $D_i\ll 1$ and $\gamma_i\ll 1$.

After realizing what are the necessary and favoring conditions for the 
transient effect of a negative--current fluctuation, let us elucidate 
sufficient conditions for the inequality $J(t)<0$. Since the latter can 
happen only at $t\ll 1$ and with a nonuniform density of carriers, let us 
take $t=0$ and the maximally nonuniform initial density
\be
\rho_i(x,0)=f_i(x) = Q_i\delta(x-a_i) ,
\ee
resulting form eq.(14) under $b_i\ra 0$. Then eq.(12) yields
\be
J(0) =\mu_1Q_1E(a_1,0) +\mu_2Q_2E(a_2,0).
\ee
And for the electric field (10) we get
\be
E(x,0) = 1 +4\pi Q_1\left [ a_1 -\Theta(a_1 -x)\right ] +
4\pi Q_2\left [ a_2 -\Theta(a_2-x)\right ] ,
\ee
where $\Theta(x)$ is the unit--step function. From eqs.(18) and (19), we 
find that the inequality $J(0)<0$ acquires the form
$$ 
\mu_1Q_1\left\{ 1 +4\pi Q_1\left ( a_1 -\frac{1}{2}\right ) +4\pi 
Q_2\left [ a_2 -\Theta(a_2 -a_1)\right ]\right\}  + $$
\be
\mu_2Q_2\left\{ 1 +4\pi Q_2\left ( a_2 -\frac{1}{2}\right ) +4\pi 
Q_1\left [ a_1 -\Theta(a_1 -a_2)\right ]\right\}  < 0 .
\ee
Inequality (20) is a sufficient condition for the occurrence of the 
transient effect of a negative electric current.

There exist a number of cases when inequality (20) holds true. To prove 
that such cases do exist, consider a particular example when 
$a_1=a_2\equiv a$. Then, from eqs.(18) and (19), we derive that $J(0)<0$ 
means that
\be
(\mu_1Q_1+\mu_2Q_2)\left [ 1 +4\pi Q\left ( a-\frac{1}{2}\right )\right ] < 0 ,
\ee
where \be
Q\equiv Q_1+ Q_2 .
\ee
Depending on whether the charge (22) is positive or negative, we have 
from (21) either
\be
a < \frac{1}{2} -\frac{1}{4\pi Q} \qquad ( Q > 0 ) 
\ee
or, respectively,
\be
a > \frac{1}{2} +\frac{1}{4\pi|Q|} \qquad ( Q < 0) .
\ee
Remembering that $0<a<1$, we see that both eqs. (23) and (24) can be 
satisfied only when
\be
|Q| >\frac{1}{2\pi} .
\ee
Eqs.(23), (24) and (25) are sufficient conditions for the appearance of 
a negative electric current at the initial stage of the process starting 
with the nonuniform distribution (17).

To confirm the above analysis, we have solved eqs.(6) and (7) numerically 
and calculated the electric current (12) as a function of time. We accept 
the case favoring the appearance of the negative current, when $D_i\ll 1$ 
and $\gamma_i\ll 1$. The initial condition (8) is taken in the Gaussian 
form (14). The voltage integral (9) plays the role of a boundary 
condition for the electric field. For the carrier densities we may accept 
the Neumann or Dirichlet boundary conditions [1]. We considered both of 
them and found that the general behaviour of solutions is practically the 
same, but the calculational procedure is less stable in the case of 
Dirichlet conditions. Therefore, for the reason of stability, we preferred 
the Neumann boundary conditions. The numerical solution of eqs.(6) and 
(7) was accomplished using the third--order Rusanov finite--difference 
scheme [10].

The results of numerical calculations are in complete agreement with the 
necessary and sufficient conditions, obtained above, for the appearance 
of a negative electric current. Consider, e.g., the case of unipolar or 
majority--carrier devices such as metal semiconductor field--effect 
transistors, Schottky varactor diodes, and $GaAs$ transferred electron 
devices [1] that are modelled by a single species set of the basic 
transport equations. In the case of one type of charge carriers, all 
processes are absolutely similar with respect to whether this is positive 
or negative charge. Let us take $Q_1\equiv Q\neq 0$ and $Q_2=0$.
Also, for brevity, we shall write $a_1\equiv a$ and $b_1\equiv b$. The 
time behaviour of the electric current (11) is presented in Figs.1 to 5.

Fig.1 demonstrates that if condition (25) does not hold then the negative 
electric current does not appear. In Fig.2, the charge satisfies condition 
(25), and the negative current arises provided condition (23) is valid. 
Fig.3 emphasizes the necessity of an essentially nonuniform distribution 
of the carrier density at initial time. Fig.4 illustrates that a 
nonuniform initial density is necessary but not sufficient: to produce a 
negative current, the location of this nonuniformity is to satisfy 
condition (23). Fig.5 stresses the importance of the latter condition 
showing that no negative current appears, even for a large nonuniform 
initial density, if this condition does not hold true.

Now let us discuss the following two important questions: (i) How the 
studied electric current $J(t)$ can be measured? (ii) What could be 
possible physical applications of the discovered effect?

The quantity that can be measured directly, as is known [1], is the 
density of the total current across the ends of a device. That is, the 
directly measurable quatities are the density of current across the left end,
\be
J(0,t)\equiv j_{tot}(0,t) ,
\ee
and the current density across the right end,
\be
J(1,t)\equiv j_{tot}(1,t) .
\ee
How the measurable end currents (26) and (27) are related to the studied 
total current (11)?

According to the detailed analysis made above, the negative--current 
effect can arise as a transient phenomenon at the initial stage of the 
time evolution. Consider the relaxation process in the interval of time 
$0\leq t\leq t_0$, with $t_0\ll \gamma_i^{-1}$. In the case when 
$\gamma_i\ll 1$, we may have $t_0\gg 1$, so that the considered time 
interval covers all stages of interest. For this time interval, the 
evolution equations, following from (1) and (2), in the plane geometry 
studied are
$$ \frac{\prt\rho_i}{\prt t} +\frac{\prt j_i}{\prt t} = 0 , \qquad
\frac{\prt E}{\prt x} = 4\pi(\rho_1 +\rho_2) . $$
Basing on these equations, it is straightforward to derive that
\be
\frac{\prt}{\prt x} j_{tot}(x,t) = 0 ,
\ee
where $j_{tot}$ is the total current density (4). Eq.(28) tells us that 
$j_{tot}(x,t)$, actually, does not depend on the space coordinate $x$. If 
so, then it is obvious that
\be
J(t)\equiv\int_0^1 j_{tot}(x,t)dx = j_{tot}(x,t) . 
\ee 
Consequently, the end currents (26) and (27) are the same as the total 
current (29),
\be
J(0,t) = J(1,t) = J(t) .
\ee
Thus, the studied total current $J(t)$ is itself a directly measurable 
quantity.

In numerical calculations, it is admissible to study any of the currents 
in eq.(30). However, the integral representation (5) or (29) is more 
suitable for analytical investigation. This is why we dealt with this 
representation for deriving the explicit conditions under which the 
negative--current effect could arise.

It is worth emphasizing that in all our consideration the total current 
density (4) always was taken in the complete form including the displacement
current. The latter nowhere was neglected. The fact that, passing from 
eq.(5) to eq.(11), the displacement current formally disappeared in the 
integrand is due to the advantage of using the integral representation 
(5) for the total current and because of the voltage condition (9). 
Really, starting from eq.(5), we have
$$ \int_0^1 j_{tot}(x,t)dx = \int_0^1 j(x,t)dx +
\frac{\varepsilon}{4\pi}\int_0^1\frac{\prt}{\prt t} E(x,t) dx , $$
where
$$ j(x,t)\equiv j_1(x,t) +j_2(x,t) . $$
Invoking the voltage condition (9), we get
$$ \int_0^1\frac{\prt}{\prt t} E(x,t) dx =\frac{\prt}{\prt t}\int_0^1
E(x,t)dx = 0 . $$
Consequently, we obtain the identity
$$ \int_0^1 j_{tot}(x,t) dx = \int_0^1 j(x,t)dx , $$

Certainly, eqs.(5) and (11) would not be identical if the applied voltage 
would not be constant. If instead of (9)  we would have the voltage condition
$$ \int_0^1 E(x,t)dx = \varphi(t) , $$
with a time--dependent voltage $\varphi(t)$ in dimensionless units, then 
instead of (11) we would get
$$ J(t) =\int_0^1 j(x,t)dx +\frac{\varepsilon}{4\pi}\frac{d}{dt}\varphi(t) . $$
The fact that under an oscillating applied voltage electric current 
oscillates as well can surprise nobody. This is why we concentrated on 
the case of a constant applied voltage. In the latter case, strong 
fluctuations of electric current are not expected. And the possibility that
electric current turns against the applied voltage seems surprising.

The discovered unusual effect of negative electric current is of interest by 
itself, irrespectively of its feasible applications. Nobody can guarantee 
in advance what could be the most important use of a new effect. 
Nevertheless, we may suggest some possible physical applications.

Since the appearance of the negative electric current is closely related 
to the properties of charge carriers in semiconductor, one can, by 
measuring this current, extract information about the characteristics of the
carriers, for example, about their mobilities.

Another possible application of the effect has to do with the case when 
the initial charge distribution is formed by irradiating a semiconductor 
sample by ions. In such a case, the charge distribution is a Gaussian 
centered at the distance, from the irradiated surface, of a mean free 
path of ions. The occurrence of the negative electric current essentially 
depends, according to eqs.(23) and (24), on the location of the initial 
charge distribution. Therefore, the negative--current effect can be used for
measuring the mean free paths of ions in irradiating beams.

One more application is related to the fact that semiconductor devices work
very often under the influence of radiation. For instance, such devices 
can be a part of electronic systems serving near atomic stations or in 
cosmic space. The influence of radiation can lead to the development of 
nonuniformities spoiling the work of semiconductor devices. This, in 
turn, can result in misfunctioning of electronic systems, which can 
provoke dramatic consequences. Before this happens, it would be desirable 
to get a warning that nonuniformities due to irradiation have reached the 
dangerous level. Then one could prevent dramatic accidents. As far as a 
growing nonuniformity in a device can yield the occurrence of negative 
electric current, this effect could be used as a kind of warning signaling 
that the situation is getting dangerous. If inside an electronic system 
acting under the influence of irradiation one incorporates specially tuned 
semiconductor devices, based on the effect of negative current, then such 
devices can play the role of controllers.

It is possible to speculate more suggesting other ways of using the 
effect. However, we would like to stress it again that a new physical 
phenomenon is of value as such. All experience teaches us that the most 
appropriate usage of a newly discovered effect practically never can be 
predicted in advance.

In conclusion, we have shown that in nonequilibrium semiconductors a 
negative electric current can arise, being directed against the applied 
voltage. This effect is principally transient and can appear only at the 
initial stage of the process, due to a nonuniform initial carrier 
density. The necessary and sufficient conditions for the occurrence of 
this effect are derived. The solution of the transport equations has been 
done both analytically and numerically.

\vspace{5mm}

{\bf Acknowledgement}

\vspace{3mm}

A grant from the National Science and Technology Development Council of 
Brazil is appreciated.

\newpage

\begin{center}
{\bf Figure Captions}
\end{center}

{\bf Fig.1}

The total electric current as a function of time for the initial carrier 
density with $Q=1/2\pi$ and $b=0.05$, at two different locations: 
$a=0.25$ (curve 1) and $a=0.5$ (curve 2). All quantities here and in 
other figures are given in dimensionless units explained in the text.

\vspace{1cm}

{\bf Fig.2}

The total electric current as a function of time for $Q=0.5$ and 
$b=0.05$, at different initial locations of carriers: $a=0.1$ (curve 1),
$a=0.25$ (curve 2), $a=0.5$ (curve 3) and $a=0.75$ (curve 4).

\vspace{1cm}

{\bf Fig.3}

The time dependence of the total electric current for $Q=0.5$ and $a=0.25$, 
at different widths of the initial nonuniform density of carriers: $b=0.05$ 
(curve 1), $b=0.1$ (curve 2) and $b=0.5$ (curve 3).

\vspace{1cm}

{\bf Fig.4}

The total electric current vs. time for $Q=3$ and $b=0.05$, at different 
initial carrier locations: $a=0.1$ (curve 1), $a=0.25$ (curve 2), 
$a=0.5$ (curve 3) and $a=0.75$ (curve 4).

\vspace{1cm}

{\bf Fig.5}

The total electric current vs. time for $Q=3$ and $a=0.5$, at different 
widths of the initial carrier distribution: $b=0.05$ (curve 1) and 
$b=0.5$ (curve 2).

\end{document}